# Chiral plasmonic nanocrystals for generation of hot electrons: towards polarization-sensitive photochemistry

Tianji Liu,[1,2,†] Lucas V. Besteiro,[1,3,†] Tim Liedl,[4] Miguel A. Correa-Duarte,[5], Zhiming Wang,*[1] and Alexander Govorov*[1,2]

[1]Institute of Fundamental and Frontier Sciences, University of Electronic Science and Technology of China, Chengdu 610054, China

[2]Department of Physics and Astronomy, Ohio University, Athens, Ohio 45701, United States

[3] Centre Énergie Matériaux et Télécommunications, Institut National de la Recherche Scientifique, 1650 Boul. Lionel Boulet, Varennes, QC J3X 1S2, Canada

[4]Fakultät für Physik and Center for Nanoscience, Ludwig-Maximilians-Universtät München, Geschwister-Scholl-Platz 1, 80539 Munich, Germany

[5]Department of Physical Chemistry, Center for Biomedical Research (CINBIO), Southern Galicia Institute of Health Research (IISGS), and Biomedical Research, Networking Center for Mental Health (CIBERSAM), Universidade de Vigo, 36310 Vigo, Spain

[†] Equal contributors

**Abstract:** The use of biomaterials - with techniques such as DNA-directed assembly or bio-directed synthesis - can surpass top-down fabrication techniques in creating plasmonic superstructures, in terms of spatial resolution, range of functionality and fabrication speed. Particularly, by enabling a very precise placement of nanoparticles in a bio-assembled complex or a controlled bio-directed shaping of single nanoparticles, plasmonic nanocrystals can show remarkably strong circular dichroism (CD) signals. Here we show that chiral bio-plasmonic assemblies can enable polarization-sensitive photochemistry based on the generation of energetic (hot) electrons. It is now established that hot plasmonic electrons can induce surface photochemistry or even reshape plasmonic nanocrystals. Here we show that merging chiral plasmonic nanocrystal systems and the hot-election generation effect offers unique possibilities in photochemistry - such as polarization-sensitive photochemistry promoting nonchiral molecular reactions, chiral photo-induced growth of a colloid at the atomic level and chiral photochemical destruction of chiral nanocrystals. In contrast, for chiral molecular systems the equivalent of the described effects is challenging to be observed because molecular species exhibit typically very small CD signals. Moreover, we compare our findings with traditional chiral photochemistry at the molecular level, identifying new, different regimes for chiral photochemistry with possibilities that are unique for plasmonic colloidal systems. In this study, we bring together the concept of hot-electron generation and the field of chiral colloidal plasmonics. Using chiral plasmonic nanorod complexes as a model system, we demonstrate remarkably strong CD in both optical extinction and generation rates of hot electrons. Studying the regime of steady-state excitation, we discuss the influence of geometrical and material parameters on the chiral effects involved in the generation of hot electrons. Optical chirality and the chiral hot-electron response in the nanorod dimers result from complex inter-particle interactions, which can appear in the weak coupling regime or in the form of Rabi splitting. Regarding practical applications, our study suggests interesting opportunities in polarization-sensitive photochemistry, chiral recognition or separation, and in promoting chiral crystal growth at the nanoscale.

**Introduction.** In materials with a large number of mobile electrons, such as metals, these charge carriers can be displaced by external electric fields, and we refer to the resonant modes excited in such a way as plasmons. The field of plasmonics has grown rapidly in the past decades, motivated by the capabilities that they afford us in manipulating light in the nanoscale. By constructing plasmonic systems in the nanometer scale we are, effectively, employing antennas that couple strongly with electromagnetic radiation at frequencies up to the UV spectral range [1] and localize its radiant energy. In doing so we can manipulate the propagation of light at this scale and create novel optical effects, [2–6] enhance secondary such as Raman scattering [7,8] or efficiently promote the conversion of light into other forms of energy. [9–11]

Among the topics where the use of plasmonic nanoparticles has extended our scientific purview has been the study of optical chirality. Plasmonic nanoparticles and assemblies have proven useful not only in facilitating strategies to measure the chirality of molecular analytes by enhancing and shifting their optical circular dichroism (**CD**) signal,[12] but also in creating artificial chiral systems with a strong and controllable differential optical response to left and right circularly polarized light (**CPL**). This latter aspect has been naturally developed in the creation of patterned surfaces through top-down fabrication methods,[13–17] but self-assembly fabrication techniques have also opened the door to the creation of complex colloidal plasmonic bio-assemblies with a tailored chiral response.[13,18–25]

Regarding the localization and conversion of radiant energy, plasmonic nanoparticles have been used in a variety of ways, including the enhancement of photovoltaic devices through different mechanisms, [10,14,26,27] driving photocatalytic processes [28–38] or heating in the nanoscale. [17,39–42] All of these disparate applications take advantage of different physical processes, but they have in common their exploitation of a fundamental property of plasmonic excitation, the enhancement of the electric field inside and around the plasmonic nanostructures. Here we would like to highlight in particular the application of plasmonic enhancement to act as a photocatalyst, driving chemical reactions of current scientific and technological relevance such as water splitting, [43] $H_2$ dissociation [28] or $CO_2$ reduction. [44] Two ways in which plasmonic nanoparticles can serve that purpose is by providing additional energy for the reaction by heating their surrounding medium, [17,39–42] and to generate and donate excited (hot) charge carriers to the reacting species. [28–38] The injection of hot electrons (**HE**s) as a photocatalytic process is very favorable from an energetic standpoint, because the excitation energy required to transfer an electron from the metal Fermi sea to the molecular excited state is lower than direct optical transitions in the molecule, which extends the usable spectral range of solar radiation. Furthermore, HE generation can be boosted through techniques that enhance the induced field in the plasmonic nanoparticle, such as the creation of hot spots. [45–49]

The confluence of the fields of chiral engineering and energy conversion has the potential to develop new scientific and technological opportunities, both in terms of allowing us additional control parameters over plasmonic-driven processes and outlining new detection techniques for nanoscale chirality. The current literature shows successful attempts in using planar chiral metamaterials in connection with HE injection in semiconductors, [14,15] photochemistry [16] and photoheating. [17,40]

In this manuscript we aim to extend the research on of hybrid nanomaterials into a new direction, by proposing the concept of polarization-sensitive photochemistry utilizing generation of hot electrons in chiral plasmonic nanocrystals (**NCs**). In our calculations, based on realistic models, we predict that chiral plasmonic NCs are able to create a strong CD effect for the rates of generation of HE. Since HEs are able to induce chemical reactions in a matrix [50,51] and on the surface of a NC, [32,52,53] we predict that the strong CD effect for the HE generation in our system can lead to related CD effects in a variety of photochemical reactions. Such HE-induced photochemical reactions have been observed in several plasmonic systems. [32,50–53]

As a model system for the chiral HE generation effect, we take an assembly of two plasmonic NRs, which was realized experimentally in Refs. [20,21,54]. The fabrication of such bio-plasmonic assemblies is feasible thanks to DNA origami nanofabrication technology,[20,55–57] which is presently very well developed. Another suitable model for the HE generation effect in a chiral plasmonic system is a single monolithic NC with a chiral shape; such NCs were proposed by us in Ref. [58]. Among the different examples of research published on such monolithic chiral NCs, [59–63] the NCs reported in the very recent paper [63] are really suitable for the processes described herein, since they have very large chiral asymmetry factors. The other part of the discussed concept is the generation of HEs and injecting them into small $TiO_2$ nanoparticles (or clusters) deposited on the NC surface. The electrons generated by the NC plasmon and then transferred to $TiO_2$ can in a later step participate in chemical reaction in a liquid, taking advantage of the energy alignment of $TiO_2$ conduction band and certain molecular excited states. This scheme with HEs and small $TiO_2$ nanoparticles was realized by us in Refs. [50,51]. In the papers referenced above, long-lived plasmonic electrons injected into the small titania nanoparticles were transferred to the solution creating reactive radicals that led to the photodegradation of dye molecules. On the other hand, the NCs used in Refs. [50,51] were not chiral and, therefore, the observed strong photochemistry was not be sensitive to the circular polarization of exciting light.

In this paper, we demonstrate very large CD signals in the HE generation, which can in turn induce chemical reactions. [32,50–53] In fact, we predict and propose polarization-sensitive photochemistry with very large asymmetry g-factors, in the order of 0.15-0.6 (15-60%). Such giant g-factors are only possible for chiral plasmonic systems with very strong absorption resonances and strong plasmonic near-field interactions. To compare this category of systems with chiral molecular systems, we now look at the typical g-factors of representative chiral molecules. For example, the g-factor of proteins with the α-helix secondary structure, which is strongly chiral, is ~ $10^{-3}$ [64]; some other molecules may have g-factors like ~ 0.05 (polyaromatic compounds from Ref. [65]) or ~0.01 (alleno–acetylenic macrocycles from Ref. [66]).

It is interesting to compare our photophysical and photochemical mechanisms with those of traditional chiral photochemistry, which deals directly with chiral molecules. [67] The principle difference between our proposed effects and traditional chiral photochemistry is that the latter studies asymmetric photoreactions of chiral molecular species under CPL. But the mechanisms under discussion here are asymmetric in their sensitivity to CPL, not in the selectivity of molecular enantiomers. Therefore, the reacting molecules are either nonchiral or preserve their enantiomeric ratios, but the plasmonic catalyst (chiral NCs) is chiral and, moreover, exhibits unusually-large asymmetric responses to CPL. Such strong optical asymmetry in the absorption of light between the two enantiomers of chiral species is fundamentally impossible (chiral molecules can be too small in size and are just excitonic, with polarizabilities much weaker than those for a plasmonic material). In chiral photochemistry, the central characteristic is the preferential promotion of reactions with the handedness of incident light, and one can see cases in which: (1) initial molecular states are chiral; (2) products are chiral; (3) both initial and final molecular states are chiral. Products of chiral photochemistry can be nonchiral as a result of photo-destruction. An excellent review on chiral photochemistry[67] defines and discusses three groups of photochemical reactions: *Photoderacemization*, *asymmetric photodestruction, and asymmetric synthesis.* These mechanisms, which names should be self-explanatory, are of great interest for their clear applications in pharmaceutical, and overall biological applications. The origin of chiral photo reactions lies, of course, in the optical CD of chiral molecules, that is asymmetric absorption. But the molecular CD is very small, and the asymmetry of photoreaction products obtained from a racemic mixture, so-called enantiomeric excess (ee) would not be observable directly after the absorption process. To see nonzero values of ee after such a process, one should apply autocatalytic methods or other type of amplification, which are non-linear kinetic methods in chemistry. In other words, chiral photochemistry can be only based on the fundamental process of optical absorption asymmetry (i.e. CD), but should also involve non-trivial chemical processes together with the CPL illumination to produce measurable results. To give an example, the ee produced by some chiral molecular reaction can be as high as 4% (the 4-cyclo-octenone system from refs. [67,68]), which is higher than the CD for the same molecular species. Since this paper is focused on the optoelectronic and photochemical properties of chiral plasmonic NCs, providing a detailed description of chiral molecular photochemistry lies beyond its scope, and would suggest the interested reader to look into the excellent review in Ref. [67].

We now resume the discussion on the subject of this paper, which is the asymmetric generation of HEs in chiral NCs under CPL excitation. One can see the following applications for the photochemical response of chiral NCs: Case 1: *Polarization-sensitive photochemistry with nonchiral molecules.* In this case, the catalyst (i.e. the NC) is chiral and exhibits very large g-factors for both CD and the HE rates. We can start with a solution containing one enantiomer of the chiral NC, and this NC solution should have strong CD. Under CPL illumination, the HEs induce nonchiral chemical reaction at the surface or in solution; one well-documented effect of this HE injection is the photodegradation of dye molecules in solution. [50,51] Chirality of the dye molecules and their products is not relevant in this case, as the photocatalytic process will not carry its chiral asymmetry to the molecular reaction. The solution itself can be racemic, but the NC system should have only one enantiomer. Then, the photoproduction of molecules (products of degradation of dye) will strongly depend on the polarization of CPL, although the molecular ee (such ee can also be zero) will not depend on it. Calculated g-factors of such non-chiral reactions can be giant, of up to ~ 60%, as shown below. Such polarization-sensitive photochemistry can be used as an alternative, indirect method to observe chirality of nanoscale objects when usual CD spectroscopy cannot be used; for example, our method can be used when a solution is fully opaque and the optical transmission is practically zero. Case 2: *Chiral growth of nanocrystals, or "chiral photochemistry" for the chiral plasmonic nanocrystals.* It has been established that NC growth can be induced by HEs. [32,52,53,69] The rate of growth should depend on the total number of HEs generated in a complex. Then, a NC with a given chirality will grow differently under LCP and RCP light. Or, in a racemic mixture of NCs, NCs with opposite chirality can grow differently and the system will not remain racemic.[70] In fact, this effect permits chiral photochemistry at the level of a crystal structure of a colloidal complex-shape nanocrystal, and we think that this regime has not obvious analogs in chiral photochemistry with molecules. In this paper, we are showing that CD for HE-induced photochemistry can be remarkably strong for chiral plasmonic objects. For nonchiral NCs, photochemical mechanisms need a special consideration.[71] Case 3: *Asymmetric photodestruction of chiral plasmonic nanocrystals and complexes*. This application shares some similarity with *asymmetric photodestruction* for chiral molecules, but, in our case, this mechanism is applied to a colloidal NC system. It entails the selective asymmetric photo-destruction of one of the enantiomers of the plasmonic chiral complex. Under intensive excitation, HEs can dissociate a bio-assembled NR-NR complex into single NRs; in the case of single monolithic chiral NCs, the selective asymmetric destruction can occur through intense surface chemistry. The rates of HE generation in the linear plasmonic regime are: $Rate_\alpha = a_\alpha I_\alpha$, where $\alpha$ can be RCP or LCP, $a_\alpha$ is the chiral-electronic coefficient, and $I_\alpha$ is the intensity of incident CPL radiation; RCP and LCP stand for the right- and left-circular polarizations, respectively. In a chiral plasmonic NC or complex, $a_{LCD} - a_{RCP} \neq 0$ and its CD is very strong, with the calculated g-factors up to 60%. In the experiments with single monolithic NCs, the observed g-factors can be up to 20%.[63] The probability for dissociation of a chiral NR-NR complex bound via the DNA linkers certainly increases with the HE rates. Therefore, if we start with a racemic mixture of NR-NR complexes (L-pair and R-pair), a strong CPL excitation should after some time create unequal concentrations of the chiral states, mixed with non-chiral products (single NRs). This scheme resembles the typical photocatalytic experiments with chiral molecules.[67]

This paper is organized in the following way: the first section describes the model system and defines the CD values and related g-factors for optical and HE responses; the following section presents results for the chiroptical and photo-electronic properties of the NR-NR complexes; in the next section, we look at a design for a broadband chiral complex; finally, we conclude this study by summarizing the data and proposed effects.

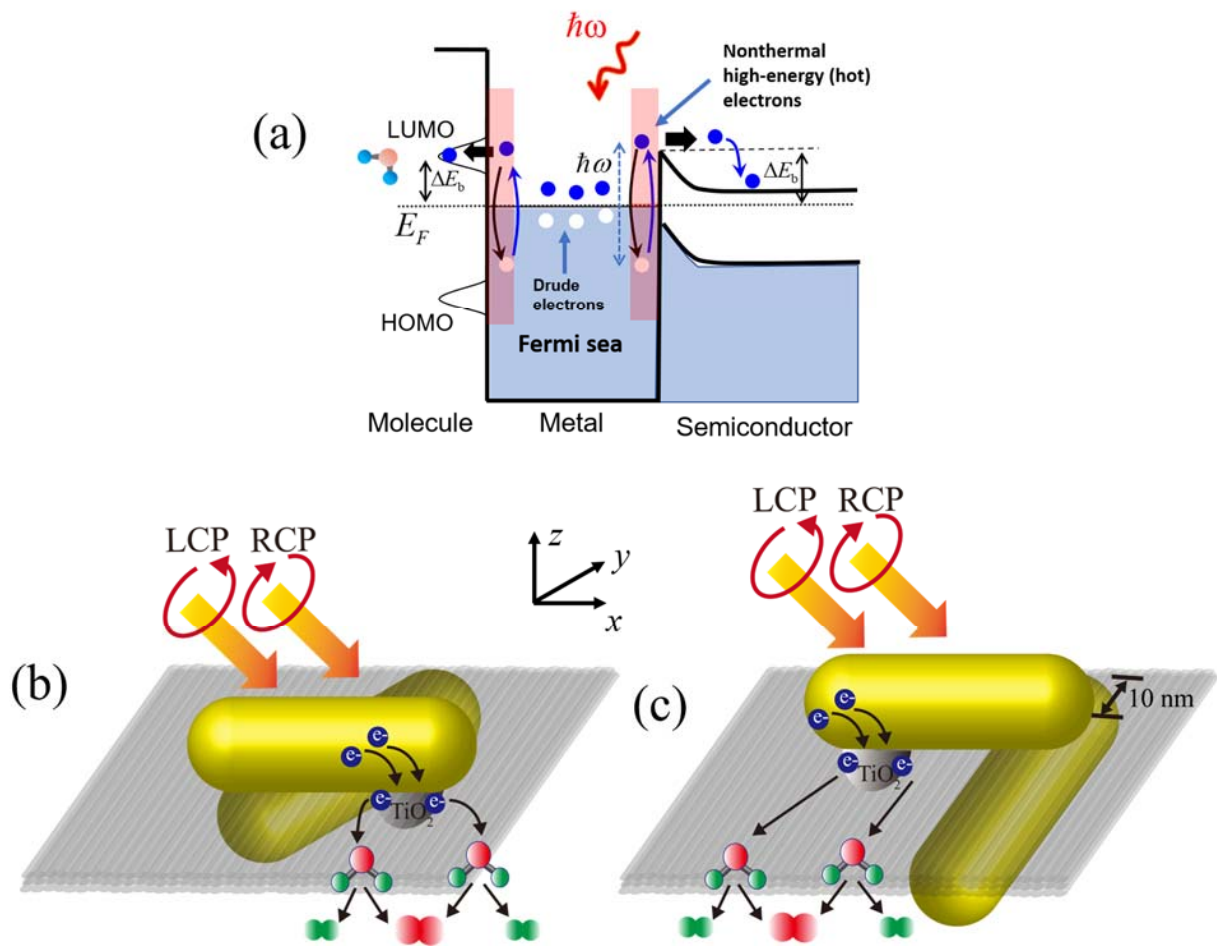

**Figure 1.** (a) Mechanisms of generation and injection of hot carriers at the interfaces of metal/semiconductor and metal/molecules, including the processes of photoexcitation and relaxation of carriers. Energetic (hot) electrons are generated near the nanocrystal surface and injected into a semiconductor material or into adsorbed chiral molecules. (b,c) Models of chiral plasmonic NRs dimers with 45° and 90° rotation angles under incident LCP and RCP illuminations. The NR sizes are: 12nm × 40nm. The DNA templates are depicted as the arrays of tubes (light grey); the vertical gap between two NRs in this model equals to the thickness of the DNA template.

In (b,c), hot electrons are first injected into small $TiO_2$ nanocrystals on the gold surface (as realized in the experimental study in Ref.[50]), that later become transferred to the solution. During this process, the charge carriers trigger the creation of reactive species (radicals) that are able to destroy dye molecules (or catalyze some other reactions) which should be present in the solution. These dye molecules can be used for an optical readout, through a process proposed and realized in Ref. [50].

**Theoretical framework.** Here we present an outline of the theory that supports our results and discussion. While we separate its description in two parts, naturally following the formal distinction between the fundamental elements in our study -chirality and hot electrons-, we also articulate their connection in our model. On one hand, we describe the main concepts regarding the optical response of chiral plasmonic assemblies, which is in turn built upon classical electrodynamics. On the other hand, we detail the fundamentally quantum mechanical phenomenon of hot electron generation in plasmonic particles.

**Chiroptical response of a plasmonic system.** The extinction cross section, $\sigma_{ext}$, of a particle or assembly quantifies the strength of its coupling with incident light, and it can be divided into two components that account for two different types of light-matter interaction: absorption cross section, $\sigma_{abs}$, and scattering cross section, $\sigma_{scat}$. Nanostructures made by noble metals (e.g. Au, Ag) are notable examples of plasmonic systems, which exhibit interaction cross sections much larger than their actual geometrical equivalents when illuminated at their resonant frequencies. Notably, when the characteristic size of a nanostructure is much smaller than the operating wavelength, i.e. in a regime well described by the quasi-static approximation, $\sigma_{scat}$ can be negligible when compared with $\sigma_{abs}$, yielding $\sigma_{ext} = \sigma_{abs} + \sigma_{scat} \approx \sigma_{abs}$. Here, $\sigma_{abs}$ is defined by the ratio of the power absorbed by the system, $Q_{abs}$, and the incident energy flux, $I_0$:

$$\sigma_{abs} = \frac{Q_{abs}}{I_0}, \quad (1)$$

$$Q_{abs} = \text{Im}(\varepsilon_{metal}) \frac{\omega}{8\pi} \int dV \, \mathbf{E}_\omega \cdot \mathbf{E}_\omega^*,$$

$$I_0 = \frac{c_0 \varepsilon_{med}^{1/2}}{8\pi} |E_0|^2,$$

where $\varepsilon_{metal}$ and $\varepsilon_{med}$ are the dielectric constant of metal and environment (medium), respectively, ω is the angular frequency of the incident light and $c_0$ is the speed of light in vacuum. The vector $\mathbf{E}_\omega$ is the complex electric field inside the metal and $|E_0|$ denotes the field amplitude of the incident light. The numerical results presented herein have been obtained using a relatively small value for the incident flux, $I_0 = 3.6 \times 10^3$ W/cm$^2$.

Two magnitudes stand at the center of the study of chiral optical activity, circular dichroism (CD) and g-factor ($g_{CD}$), which are defined in terms of the optical cross sections of a system:[17,72]

$$\text{CD} = \Delta\sigma_{ext} = \sigma_{ext,L} - \sigma_{ext,R} \quad (2)$$

$$g_{CD} = \frac{\sigma_{ext,L} - \sigma_{ext,R}}{(\sigma_{ext,L} + \sigma_{ext,R})/2}$$

where the subscripts L and R stand in place of the abbreviations LCP and RCP, respectively. One should note that, when considering an ensemble of colloidal particles, the extinctions used above should be found via the following averaging procedure:

$$\sigma_{ext,\alpha} = \langle \sigma_{ext} \rangle_{\mathbf{k},\alpha} = \frac{\sigma_{ext,\alpha,\mathbf{k}\|x} + \sigma_{ext,\alpha,\mathbf{k}\|y} + \sigma_{ext,\alpha,\mathbf{k}\|z}}{3}, \quad \alpha = L, R \qquad (3)$$

Here $\mathbf{k}\|x$, $\mathbf{k}\|y$ and $\mathbf{k}\|z$ explicitly denote the use of incident wavevectors aligned with the x, y and z directions, respectively. In other words, the values for the extinction spectra are obtained from averaging over the three main incident directions.[58,73] This treatment accounts for the random orientations of individual plasmonic dimers within the ensemble in solution. This averaging procedure is exact for nanocrystal complexes with relatively small dimensions.[58,74] It is worth noting that in the field of molecular spectroscopy the molar extinction coefficient $\varepsilon_{molar}$ is more commonly used magnitude than extinction cross sections. Thus, let us briefly mention that the conversion between $\sigma_{ext}$ (units of m$^{-2}$) and $\varepsilon_{molar}$ (units of M$^{-1}$cm$^{-1}$) is the following: $\varepsilon_{molar} = N_A \sigma_{ext} / 0.2303$, here $N_A$ is Avogadro's number.

**Generation of hot electrons in chiral systems:** The photoexcitation process entails the absorption of a photon and the connected transition of an electron to an excited state (Figure 1a). During this process, two different types of excited electrons emerge in the system[75]: one is labeled "thermalized", and their population is described by a Fermi-Dirac distribution with an effective temperature $T_e$ which is higher than that of the lattice temperature. The other type of excited electrons is termed "non-thermalized", and their population cannot be described by the Fermi-Dirac distribution, as it extends broadly to energies well above the Fermi energy ($E_F$), and up to $E_\mathrm{F} + \hbar\omega$. Such property of nonthermal, or hot, electrons is the key to use them in photocatalytic processes, as they can be transferred to suitably aligned electronic states of semiconductors and molecules that would not be energetically accessible by an internal promotion of energy $\hbar\omega$, as Fig. 1a depicts schematically. In such a case, however, the energy threshold is set so that the excited electron can overcome the energy barrier $\Delta E_\mathrm{b}$ between the materials. In plasmonic nanostructures, the transition of electrons from the Fermi sea is a collective effect, resulting in a spatial stepwise distribution of nonthermal HEs with high kinetic energy (the pink areas in Figure 1a highlight the regions, near to the NC surfaces, that will preferentially excite electrons well above the Fermi level, with energies $\varepsilon$ such that $E_\mathrm{F} + k_\mathrm{B} T_\mathrm{e} < \varepsilon < E_\mathrm{F} + \hbar\omega$). Likewise, hot holes (HHs) undergo the same process of excitation and a fraction of them will have energies such that $E_\mathrm{F} - \hbar\omega < \varepsilon < E_\mathrm{F} - k_\mathrm{B} T_\mathrm{e}$ (also, preferentially in the red areas in Figure 1a), naturally respecting the energy and charge conservation. If we couple a plasmonic nanostructure to a molecule or a semiconductor (e.g. TiO$_2$), the generated HEs can be injected into the adjacent material if the electron energy exceeds the barrier height, i.e. if $\varepsilon - E_\mathrm{F} > \Delta E_\mathrm{b}$, where they can be used for subsequent chemical reactions in the surrounding medium. It is relevant to note that the generation of nonthermal hot carriers is an ultrafast and nonequilibrium process, leading to a fast relaxation process via electron-electron (e-e) scattering, with the typical lifetime of hot carriers being shorter than 100 fs.[47,76] Therefore, most hot carriers generated near the surface of a plasmonic nanostructure will be injected into the surrounding media. At the same time, under constant illumination the system can generate hot carriers in a steady-state regime, offering a constant stream of carriers to drive chemical reactions occurring at timescales much larger than the typical lifetime of hot carriers inside the metal.

Now, to study the dependence of HE generation in the bio-assembled plasmonic complexes we need to quantify the rates of HE generation. We can derive them by solving the quantum master equation in a perturbative approach,[75,77–79] but let us just remind here the final expression for these generation rates:

$$Rate_{HE} = \frac{2}{\pi^2} \frac{e^2 E_F^2}{\hbar} \frac{1}{(\hbar\omega)^3} \int_S |E_{normal}|^2 ds, \qquad (4)$$

where $E_{normal} = \mathbf{E}_\omega \cdot \hat{\mathbf{n}}$ is the electric field internal to the nanostructure and normal to its surface, as shown in Figure 1b. In Eq. 4, $S$ is the surface of the metal structure. Importantly, Eq. 4 highlights that the excitation of HE is a surface quantum effect, and it originates from the non-conservation of the electron's linear momentum due to its scattering by the surface. In contrast, classical electromagnetic absorption is a volume effect associated with the Ohmic heat (Eqs. 1).

When incident photons carry an energy $\hbar\omega > \Delta E_b$, the corresponding excited fraction of the distribution of HEs can effectively overcome the barrier between the metal and the adsorbed molecules or semiconductor (see Figures 1b,c). The generation rates of such high-energy HEs ($Rate_{high\ energy}$) can be calculated as: [75,77–79]

$$Rate_{high\ energy} = \frac{2}{\pi^2} \frac{e^2 E_F^2}{\hbar} \frac{(\hbar\omega - \Delta E_b)}{(\hbar\omega)^4} \int_S |E_{normal}|^2 ds \qquad (5)$$

Obviously, $Rate_{high\ energy}$ includes only a fraction of $Rate_{HE}$, for electrons with energies beyond a given threshold, i.e. $Rate_{high\ energy} < Rate_{HE}$. Note that $Rate_{high\ energy}$ is usually not equal to the injection rates of HEs ($Rate_{injection}$) but it is reasonable to assume that the latter will be proportional to the former, so that an increase in high energy HE generation will lead to a proportional increase in HE injection. In practice, $Rate_{injection}$ is influenced by various factors, such as reflection at the interface and a tunneling effect.[36]

Since these rates depend on the intensity of the electric field at the surface of the particle, they will of course depend on the direction of incidence of light. Thus, as we deal with a system in solution, we again need to average this physical quantity:

$$Rate_{\text{high energy},\alpha} = \left\langle Rate_{\text{high energy}} \right\rangle_{\mathbf{k},\alpha} = \frac{Rate_{\text{high energy},\alpha,\mathbf{k}\|x} + Rate_{\text{high energy},\alpha,\mathbf{k}\|y} + Rate_{\text{high energy},\alpha,\mathbf{k}\|z}}{3}, \quad \alpha = L, R$$

(6)

Finally, to quantify the chiral response in terms of HE generation, itself a proxy for photocatalytic reaction rates, we should define a novel extension to the aforementioned optical chiral parameters: hot-electron CD and its related g-factor,

$$CD_{\text{high energy}} = Rate_{\text{high energy},L} - Rate_{\text{high energy},R} \qquad (7)$$

$$g_{CD,\text{high energy}} = \frac{Rate_{\text{high energy},L} - Rate_{\text{high energy},R}}{(Rate_{\text{high energy},L} + Rate_{\text{high energy},R})/2},$$

These new variables are responsible for characterizing the polarization-sensitive generation of HEs and the related photochemical CD.

Regarding specific implementations of chiral photocatalysts, let us briefly comment on the viability of creating such systems and on their use in conjunction with $TiO_2$ nanoparticles/clusters to realize HE photo-chemical experiments, in a manner similar to experiments conducted with achiral photocatalysts. [50,51] This computational paper studies as a model system a chiral NR-NR assembly which has already been experimentally demonstrated. [20,21,54] Other systems suitable for use as chiral photocatalysts have been realized experimentally, such as monolithic chiral NCs with a cubic frame [63] which does not require the bio-assembly of separate achiral NCs. Simultaneously, strong photo-chemistry based on plasmonic hot electrons has been demonstrated for achiral Au NCs covered with $TiO_2$ nanoparticles. [50,51] To activate our chiral NCs photo-chemically, one can grow small $TiO_2$ nanoparticles on the Au surface replacing some of the DNA ligands. One can also deposit the chiral NR-NR dimers on silica spheres prior to the $TiO_2$ growth, to make them more stable. This strategy has already been successfully demonstrated in photo-chemical experiments, where Refs. [50,51] report the use of such $TiO_2$-Au-silica hybrids. However, as a final remark on this consideration of the co-catalyst for the plasmonic system we note that, for the asymmetric photo-destruction of chiral NR-NR pairs, it wouldn't be necessary to use any co-catalytic $TiO_2$ clusters.

**Results and discussions.** The model system under consideration is illustrated in Figures 1b,c, which show two different chiral plasmonic dimers with relative rotation angles between the NRs of 45° and 90°, respectively. Hereafter, we will refer to the systems by using these angles as labels. Such well-defined systems can be straightforwardly fabricated using the DNA origami nanofabrication technique. [20,21] In our model system, the dimers are made Ag and Au, materials representative of noble metals with strong plasmonic responses. The sizes of the individual NR are taken to be 12 nm × 40 nm, much smaller than the operating wavelength of the system, ranging from 400 to 900 nm. The gap between the two NRs is varied within the interval 5-90 nm, accounting for the capability of experimentally controlling this distance by changing the number of DNA layers of origami template. For the 45° dimer, the centers of the two NRs overlap in the projected x-y plane (Figures 1b,2). For the 90° case, their centers are shifted, and they instead overlap closer to their corners, as shown in Figure 1c,2, in order to produce a chiral geometry. These geometrical parameters correspond to those of existing experimental reports.[20,56] The complex permittivity of Ag and Au are taken from the literature,[80] and the solvent is water ($\varepsilon_{med} = 1.8$). In order to simplify the computational approach and to provide a clearer, more general discussion of the phenomenon, we consider the generation of HEs in metallic structures immersed in a homogeneous dielectric medium, neglecting the screening effect of very small intermediate $TiO_2$ nanocrystals. Furthermore, although our description does not include explicit modelling of the carrier transfer through the $TiO_2$ towards the molecular compounds, this simplified description remains informative of the general picture because the above screening effect can only cause a small red shift of the plasmonic peaks.[50]

As an initial step towards understanding the differential chiral performance of HE generation in these systems, we begin by discussing the chiral optical profiles of the chiral plasmonic dimers. Using Eqs. 1,2 with data obtained from classical electrodynamic simulations (full details can be found in the SI), we calculated the optical chiral performance of the two types of plasmonic dimers with different gap sizes (5 and 13 nm). Results for Au can be found in Fig. 2, while those for Ag are in the SI, and they depict the full optical profile of the systems under the two polarizations of CPL, as well as the spectra of the differential chiral variables, CD and g-factor. As a first observation, we can point out to Fig. 2d (and Fig. S2d), which showcase values of g-factor exceeding 0.4 for the 45° Au dimer (above 0.6 for the 45° Ag dimer). These are orders of magnitude larger than those for molecular systems, with g~$10^{-3(-4)}$,[64,81] and arise from the strongly chiral symmetry of the engineered plasmonic assemblies. Now, if we look closely at the data in Fig. 2, we will see that the illumination of our system by LCP and RCP light (Figs. 2a,b,c,f) excites strong longitudinal plasmonic resonances in the interval 700-800 nm. This is accordance to what one would expect from Au NRs of such aspect ratio immersed in water. Importantly, we observe in our system a large Rabi-splitting of the individual NR plasmonic modes for small gaps, which is induced by the near-field NR-NR interaction. We also see an absorption band in the interval 400-550 nm, which comes from the transverse plasmons in the NRs and from the interband transitions in gold.

At this point, it is interesting to discuss in some more detail the physical picture of the plasmonic modes and the Rabi-splitting effect, before delving into the discussion of the HE generation in these systems. We can better see the underlying physics of the plasmonic modes by considering the maps of free surface charge density for the 45° system at the top of Figure 2. In them, the different characters of the chiral modes are apparent: the symmetrical charge distribution at $\lambda_0 = 710$ nm has a strongly dipolar character, while the asymmetrical charge distribution at $\lambda_0$=790 nm reveal a more complicated multipolar mode. Correspondingly, the higher energy mode couples more strongly with the far-field than the lower energy, multipolar one (see small gap curves in Figs. 2a,b, 3a,b, S2a,b, S3a,b). Such distributions correspond to the two normal modes in the small-gap system and arise from the near-field interaction of the strong dipolar plasmons of the two constituting NRs. As such, the Rabi splitting effect appears only for sufficiently small gaps between the NRs, so that it is not appreciable for our dimers with a 13 nm gap but becomes clear for the system with 5nm of interparticle distance. Furthermore, we make an interesting observation in comparing the data in Figures 2,3,S2,S3 between the Rabi-split spectra for the 45° and 90° dimers: the Rabi peaks are strongly asymmetric for the 45° NR pair but closely symmetric for the 90° complex. This can be understood in terms of the symmetry of the normal modes for each of the dimers, and it can be easily appreciated if we look at the dipolar component of these modes (Figure S4). The relevant comparison of surface charge maps and dipole diagram can be found in the Supporting Information (Figure S4), together with more details about the Rabi-splitting effect in the 45° and 90° NR dimers. To summarize, as one can see, we find very large chiral g-factors for the NR-NR geometry. In addition, we have previously demonstrated computationally [20,21] that such strong CD signals are robust against fluctuation of the geometrical angle in the NR-NR pair.

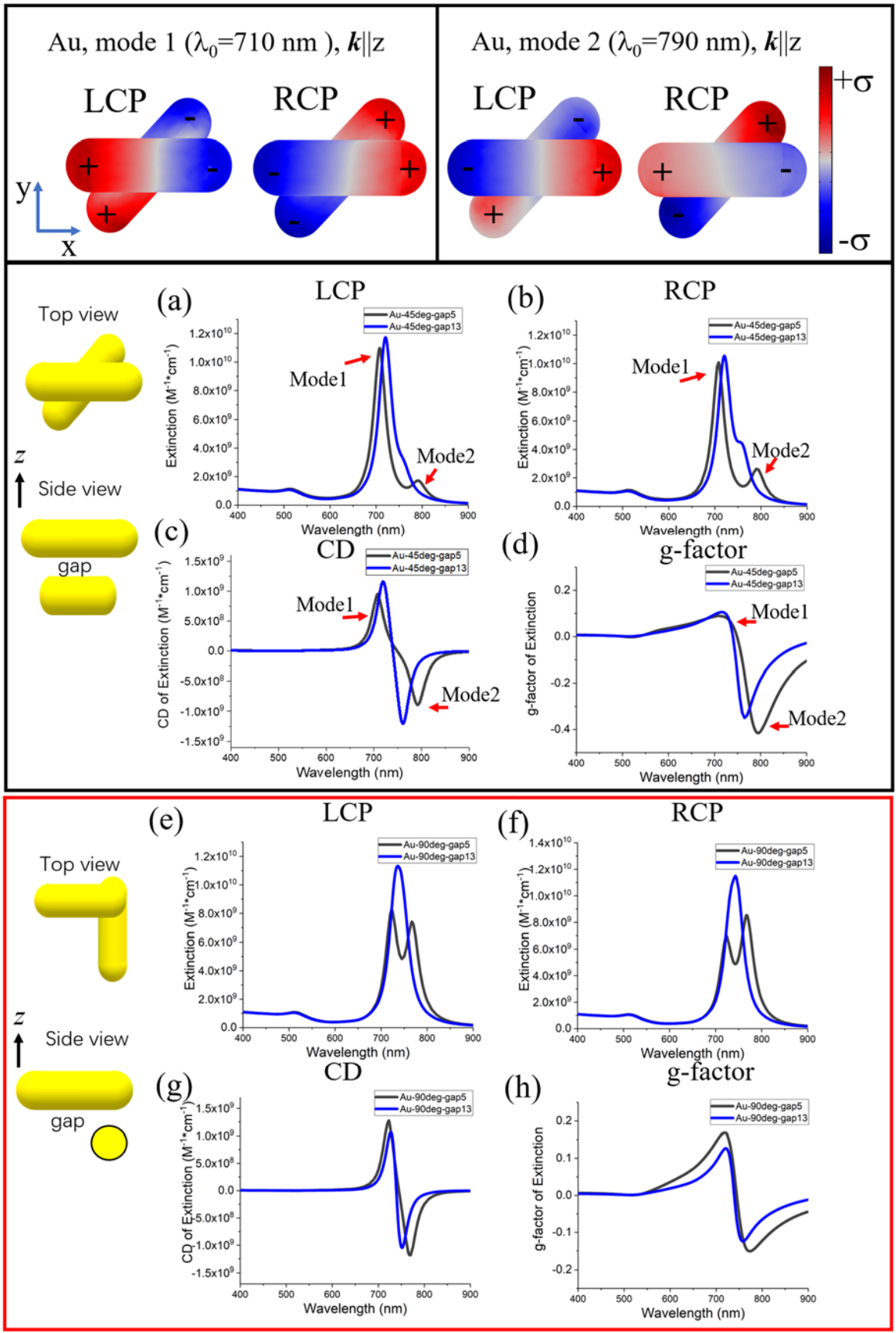

**Figure 2.** (a-b) Simulation results for the averaged extinction spectra of the NR dimers with 45° angle under CPL light with different polarizations. (c-d) Chiral optical response of the 45° dimers. These four panels show two curves each, for two different gaps between the rods. The insets on the left show the 45° dimers from two orthogonal perspectives. (e-h) The equivalent information than panels a-d, but for the 90° NR dimer. The top section of the figure shows four maps of the surface charges for the Au 45° dimer with a gap of 5 nm, at the normal modes' wavelengths and for both polarizations of light, revealing the nature of the two bright plasmonic modes in the dimers. These modes make the positive and negative chiral bands for the Au NR dimer (panels c-d).

After discussing the main features of the optical response of the dimers, we now evaluate their dynamics of HE generation by using Eqs. 6,7. The averaged generation rates of energetic HEs (*Rate* high energy) and the corresponding chiral asymmetry metrics for the different Au dimers are shown in Figure 3, whereas the equivalent data for Ag systems is in Fig. S3. It is relevant to remember that *Rate* high energy denotes the generation rate of the over-barrier nonthermal HEs with energies $\varepsilon > \Delta E_b$, which is only a fraction of the total generation rate of HEs, *Rate* HE. This barrier represents the energy threshold for the plasmonic photocatalytic system to contribute to the chemical reactions, so that we need $\hbar\omega > \Delta E_b$ in order to have some HE that can propagate into the $TiO_2$ or molecular adsorbates. At the same time, this situation is more energetically favorable than in absence of the plasmonic photocatalysts, as the energy barrier is lower than other energy thresholds in the system, such as the bandgap of $TiO_2$ or many relevant molecular transitions. For our calculations, we took $\Delta E_b$ = 1 eV as it is a representative value of a Schottky energy barrier.[45]

It is interesting to compare the optical extinction and HE spectra in Figures 2,3. In these figures, the extinction and HE spectra look similar for the spectral interval of the longitudinal plasmon resonance (650-850nm) where the intraband electronic excitations dominate and the dielectric constant is well described by the Drude model. The reason is that both effects (optical CD and HE CDs) originate from the same plasmon resonances. However, for the interband excitation interval in gold (400-600nm), the differences in the extinction-CD and HE-CD spectra should be apparent, and we indeed see that in our calculations (Figures 2,3,S5). Moreover, we note an interesting qualitative difference between the optical and photochemical CD effects. This difference comes from the fundamental underlying phenomena causing these effects, as well as from their different mathematical expressions. As mentioned above, the extinction signal for relatively small complexes, such as the ones studied here, comes mostly from the absorption inside the plasmonic components, which is a bulk effect. Correspondingly, the mathematical expressions for the optical extinction and HE generations are different. The energy dissipation in our systems is a classical effect (Drude absorption via the "frictional" Joule-heating mechanism) and occurs in a whole volume of the NRs. The case of the HE generation is different in two respects. First, it is a quantum effect (see Eqs. 4,5). Second, the HE generation is a surface effect appearing at the metal-environment interfaces due to the breaking of the linear momentum conservation for the electrons. In our NR complexes, this effect occurs preferentially at the end of the NRs, as shown in Figure 4. To conclude the comparison between optical-CD and HE-CD, we should also mention that for the Ag complexes the optical and optoelectronic spectra show different line-shapes at the main plasmonic peaks as well (Figures S2,S3). This is because the Ag NCs exhibit much stronger plasmonic enhancements and, therefore, the fine details of the physical processes start to play more important roles. Another interesting case is that of the four interacting NRs (Figure 6), where the interactions and plasmonic modes become much more complex and the physical and mathematical differences between optical-CD and HE-CD become more evident in the resulting spectra.

At this point, and using the data in Figures 3 and S3 (see Supporting Information), we can support the following conclusions: (1) Compared with the Au dimers, the Ag dimers exhibit a larger *Rate* high energy and CD in the visible wavelength interval, due to the difference between the dielectric functions of these two noble metals (see Supporting Information). (2) A remarkable Rabi splitting can be observed in all studied systems, but especially for the ones with a smaller gap size (5 nm) and for the more compact dimer (45° rotation angle); of course, this comes from the increased coupling strength between the NRs. Importantly, the Rabi splitting boosts chiral behaviors of *Rate* high energy, enhancing hot-electron g-factor of Au-dimers with 45° rotation angle (Figures 3c,d). (3) The effects of *Rate* high energy become larger in the near-infrared (NIR) interval, which is a signature of the frequency-dependent multiplier $(\hbar\omega)^{-4}$ in Eq. 5; this property looks promising for NIR photochemistry. (4) The Rabi splitting strongly influences the spectral shape of the CD signals, creating two distinct plasmon peaks (Figure 3); this effect is even stronger for the Ag complex (Figure S2,S3). (5) Overall, we observe large values of the g-factors ($g_{CD}$ and $g_{CD,high\ energy}$), with a maximum of 0.65 for the Ag dimers with 45° rotation angle and 13 nm gap, as seen in Figure S3d. Such plasmonic structures can serve as efficient chiral catalyzers in photochemistry.

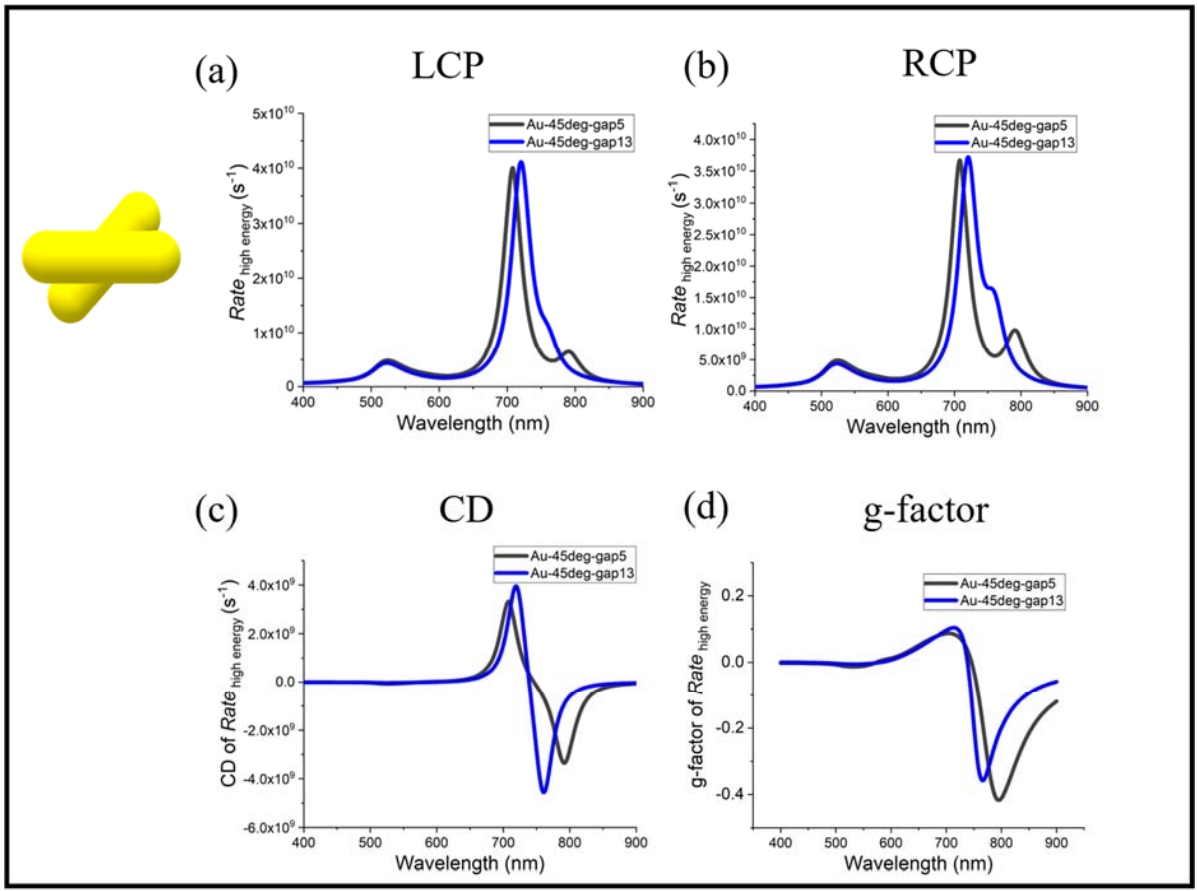
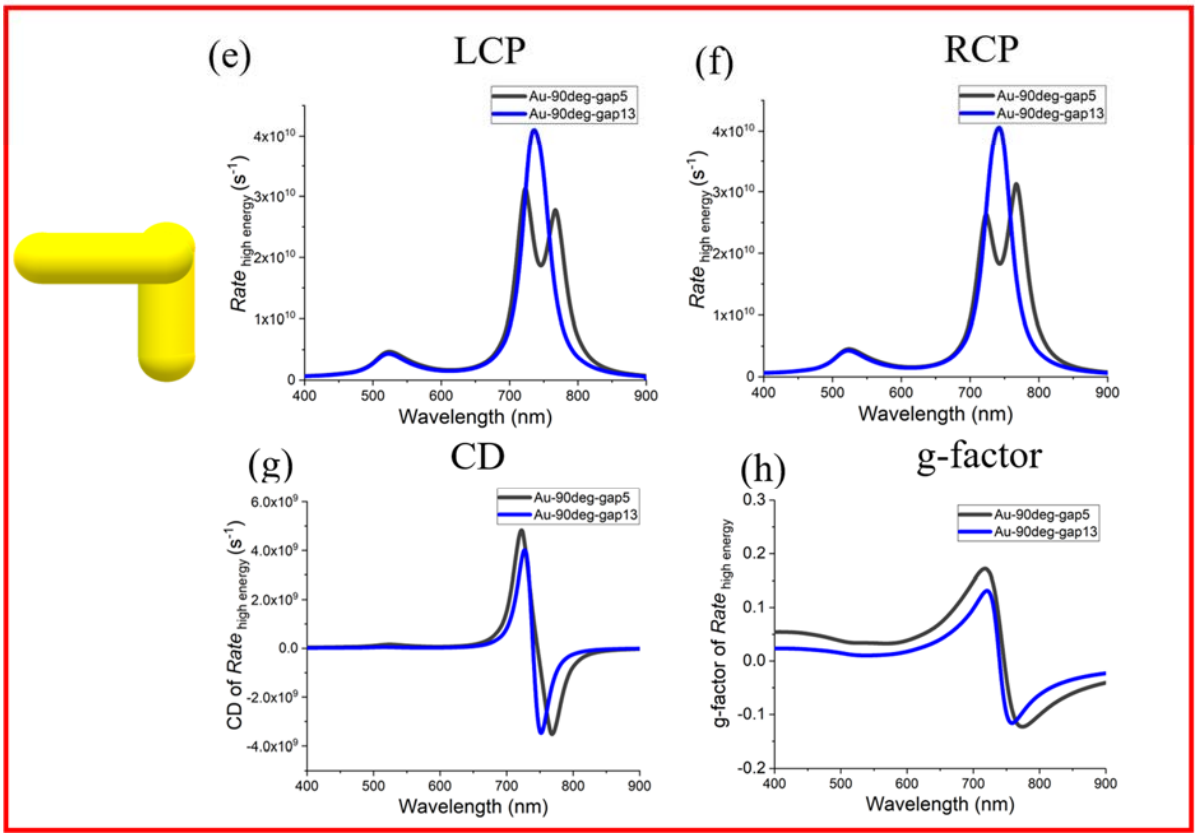

**Figure 3.** Simulation results for the averaged HE-generation spectra (a,b,e,f) and their related chiral magnitudes (c,d,g,h) for the plasmonic dimers with 45° angle (a-d) and 90° angle (e-h). Calculations were done for the Au-NRs in water for the gaps of 5 nm and 13 nm. The g-factors obtained for both systems, but specially for the 45° dimer, indicates their suitability as nanocomplexes to trigger polarization sensitive photocatalysis.

Now we take a closer look at the local maps of HE generation in Figure 4. The local generation rate of energetic HEs per surface area is defined as:[79]

$$r_{\text{high energy}}(\mathbf{r}) = \frac{2}{\pi^2} \frac{e^2 E_F^2}{\hbar} \frac{(\hbar\omega - \Delta E_b)}{(\hbar\omega)^4} |E_{\text{normal}}(\mathbf{r})|^2 \qquad (8)$$

where the position-dependent quantity $r_{\text{high energy}}(\mathbf{r})$ has the units of s$^{-1}$m$^{-2}$. In Figure 4, we show the representative maps at the characteristic wavelengths of 710 nm and 790 nm (for the Au-dimer with the 45° rotation angle and the 5 nm gap size), which correspond to the maximum (CD$_{\text{max}}$) and the minimum (CD$_{\text{min}}$) of the CD spectra in Figures 2c,3c. In these maps, we only considered incidence along the z-direction (*k*‖z) for incident circular polarized illumination, because the direction *k*‖z dominates in the averaged CD spectra. Evidently, the local surface density of generation of HEs is concentrated at the ends of NRs, resulting from the plasmonic hot spots and their enhancement of |E$_{\text{normal}}$| at plasmonic modes which fundamentally excite a combination of the dipole modes of the individual NRs (see the top of Fig. 2 and Fig. S4a). These hots spots are the small surface areas where the maximum photochemical activity is expected and, furthermore, they also have the maximum values for the local hot-electron CD (Figures 4c,f).

The above effects have relevant implications for some of the possible applications described in the introduction. The total rate of NC growth can be assumed to be proportional to the total rate of generation of HEs in a NC and, therefore, the enantiomers of a chiral NC should grow at different rates. Moreover, the local crystal growth or surface photochemistry should be proportional to the local rate of HE generation near the surface and, for a given CPL beam (RCP or LCP), surface patterns of photochemistry will be different for each of the two chiral states of the NR-NR complex. So, one particular enantiomer of the NR-NR complex will show different growth patterns and rates under different CPL illuminations (LCP or RCP). In this way, the HE CD signals can be translated to localized surface photochemistry and the related structural differences between the two configurations of the plasmonic complex.

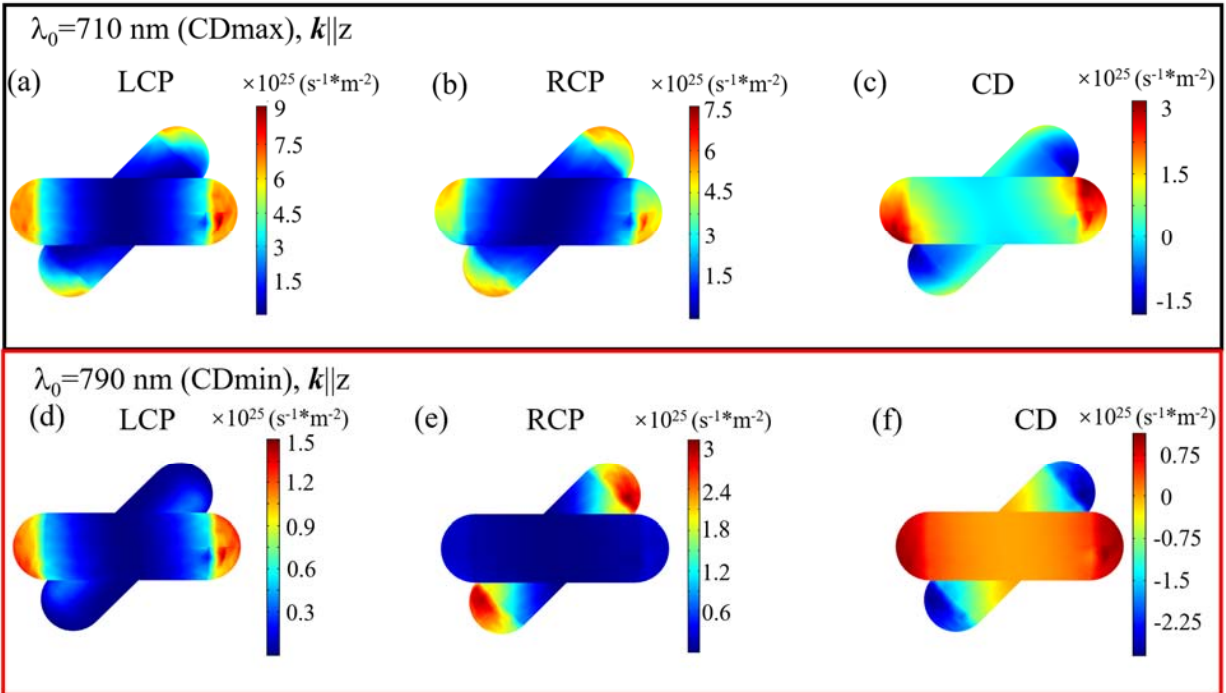

**Figure 4.** Maps of hot electron generation for the incident light propagating in the $z$-direction with wavelengths of 710 nm (CD$_{max}$) and 790 nm (CD$_{min}$). These wavelengths correspond to the two plasmonic resonances in the given dimer. Here, we show the case of the Au-dimer with 45° rotation angle and the 5 nm gap size.

Finally, we summarize the optical and HE CD effects for both geometries and for the two involved metals (Au and Ag). We observe the following features in Figure 5: (1) the HE CD mainly follows the optical CD; (2) The CDs signals for both geometries show a maximum of CD at moderate gap sizes (10-20nm). This gap size dependence can be understood physically in terms of the NR-NR interactions: at long distances the NRs start behaving as uncoupled resonators, so that the complex does not show chirality, while at very small gap sizes (<5 nm) the coupling is strong, but the collective assembly's modes begin to become closer to those of an in-plane resonator with the cross-like geometry, which is non-chiral.

**Case 1: small gaps.** For small gaps the CD signals decrease because the structure becomes geometrically less chiral; if we were to fuse the NRs making their axes overlay in one common plane, the structure would become non-chiral and the CD signals will vanish. We clearly see this trend in Figure 5 for the small gaps. Mathematically, the CD signals at small separations depend on the geometry in the following way:

$$CD \propto (k \cdot g) \cdot f(b) \qquad (9)$$

Here $k = 2\pi/\lambda$ is the wave vector of light, $g$ is the gap size and $b$ is the NR length. This functional dependence follows from the analysis made in Refs.[73,74]. The factor $k \cdot g$ comes from the retardation effects along the direction normal to both NRs (z-direction in our geometry in Figure 1), and if it becomes too small the two interacting dipoles start collapsing into a single in-plane structure.

**Case 2: Large NR-NR separations**. For large gaps, the CD decays since the dipole-dipole and electromagnetic interactions between the NRs decay with the distance. The reason is that all chiral responses in the chiral NR pairs originate from the NR-NR interactions between two single NRs that are not chiral. Asymptotically, we see from the numerical calculations that, for large $g$, $CD \propto 1/g^2$. We can understand this behavior in the following way: the dipole-dipole interaction for long distances gives the typical behavior $f(b) \propto 1/b^3$, but with the retardation coefficient $k \cdot g$ from Eq. 9, we indeed obtain $CD \propto 1/g^2$ for $g \to \infty$.

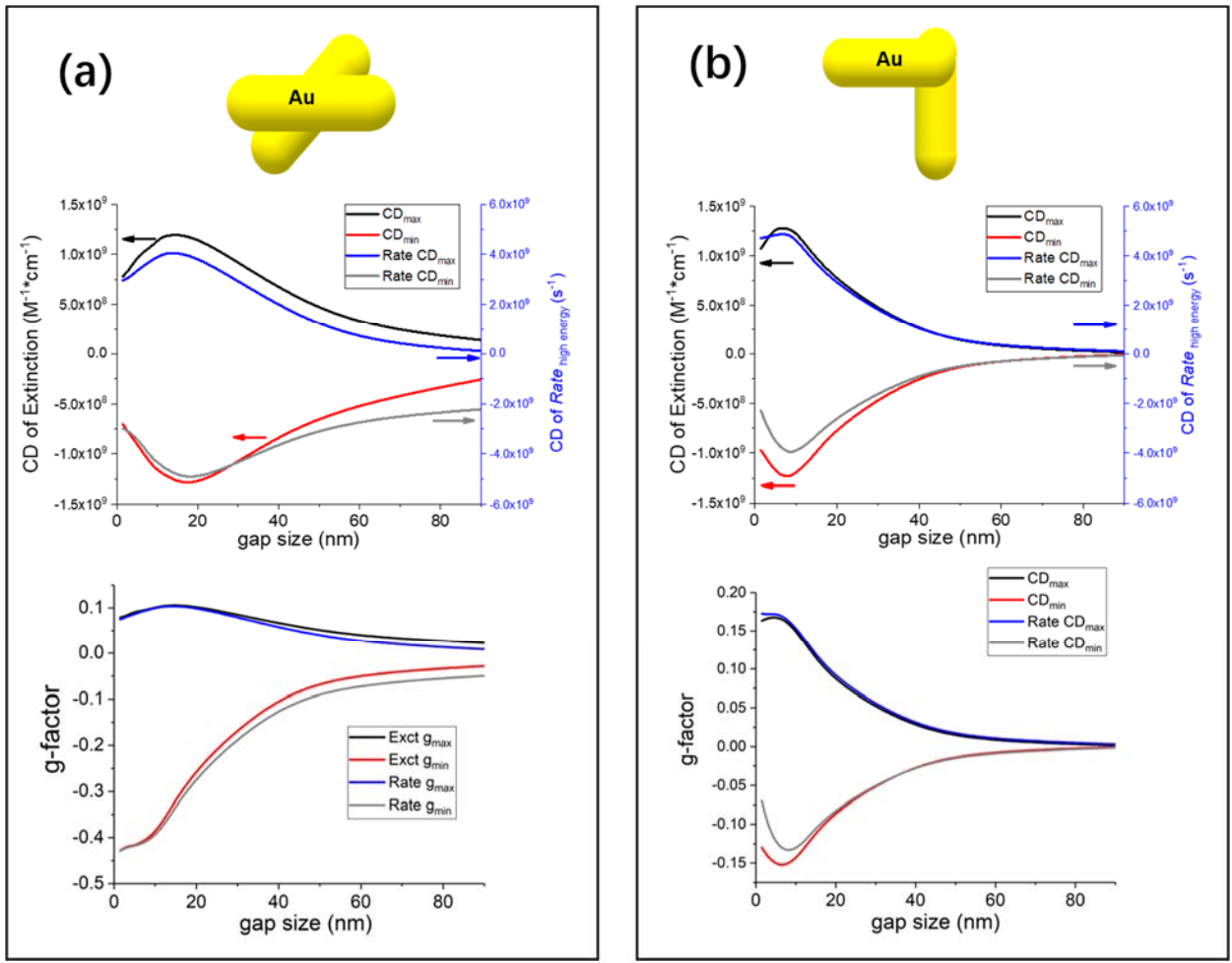

**Figure 5.** Peak values of the optical and HE chiral signals as functions of the bio-assembly gap for the 45° Au dimer (a) and the 90 ° Au dimer (b). We observe that for an increasing gap size, the CD signal decays faster for the case of the 90° dimer.

Practical applications often require, within the context of solar energy conversion, not only the efficient generation of high energy HEs at one single frequency, but also a broadband utilization of the entire solar spectrum. For this goal, we also designed a helical Au-tetramer superstructure for broadband generation of energetic (over barrier) HEs (inset of Figure 6), where the length of the AuNRs varies from 25 nm to 40 nm with a 5 nm step and their rotation angle is changed from 0° to 135° with a 45° step. The gap size is set at a constant 5 nm between each AuNR. As shown in Figure 6 a,d, the extinction and *Rate* $_{high\ energy}$ now extend from 500 nm to 800 nm, due to the different resonant frequencies of the plasmonic modes in the AuNRs. Likewise, the CD spectra and the g-factor display broader spectra for the optical and HE responses, and each plasmonic resonant mode can be recognized both in the optical and HE generation. Please note that we did not perform a geometry optimization to maximize the system's bandwidth, but took instead realistic parameters of NRs. Therefore, the performance can be further improved by optimizing geometrical parameters such as the gap size, rotation angle and length of AuNRs. Interestingly, the chiral performance is different between the optical extinction and *Rate* $_{high\ energy}$. For example, the optical CD spectrum and the extinction g-factor show a broadband bisignate line shape (Figures 6b,c). In contrast, the HE spectra (the HE CD and the g-factor for *Rate* $_{high\ energy}$) primarily show negative values (Figures 6e,f). The different chiral performances for the optical and HE chiral effects appear because these effects depend on the different components of the electric fields: the optical CD comes from the amplitude of $\mathbf{E}_\omega$ in the full volume of the particles, whereas the HE CD depends on the internal normal field near the surface, $\mathbf{E}_{normal}$. Furthermore, because of the complexity of the tetramer, this structure has more complex NR-NR interactions as compared to the NR dimer. Therefore, the CD spectra for the optical extinction and the HE effects differentiate more clearly from each other in the more complex case of the NR tetramer.

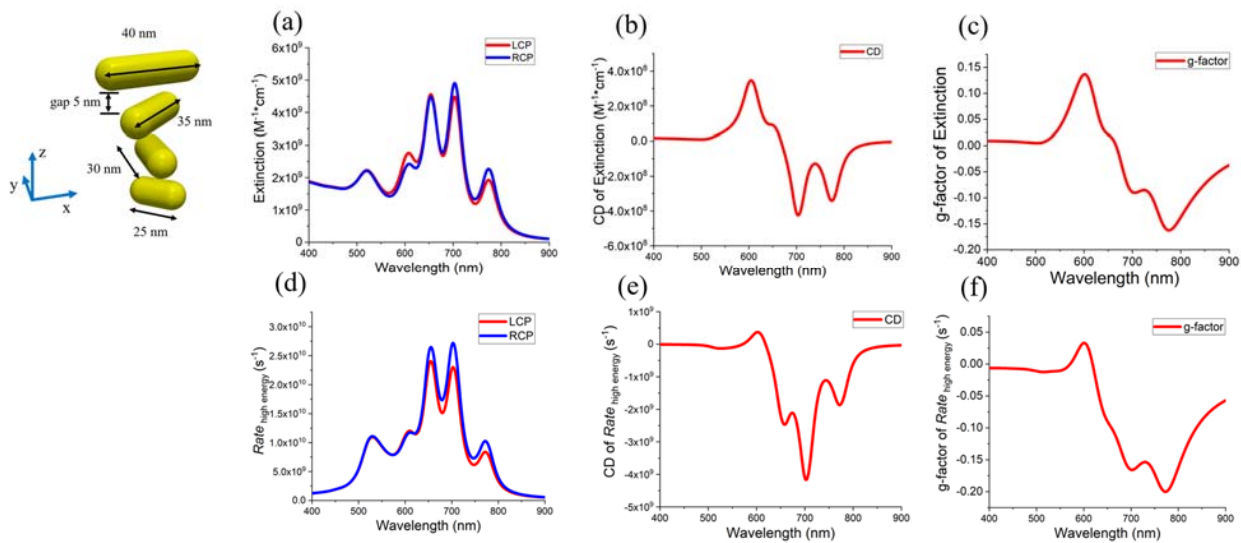

**Figure 6.** The inset shows the broadband plasmonic Au-tetramer with the step of 45° rotation angle, i.e. 0°, 45°, 90°, 135° (top-down), and 5 nm steps of the NR length, i.e. 40 nm, 35 nm, 30 nm, 25 nm (top-down), respectively. The gap between NRs is taken uniformly as 5 nm. (a-c) Extinction spectra for the circularly-polarized light, extinction CD, and g-factor. (d-f) HE spectra for the circularly-polarized light, HE CD, and the related g-factor.

In conclusion, we have demonstrated that chiral plasmonic nanostructures exhibit very large chiral optical asymmetries that become transferred into the HE generation processes. We think that this effect can be used as a new mechanism for polarization-sensitive plasmon-induced hot-electron photochemistry. The proposed chiral photochemical effect has fundamentally different properties as compared to chiral photochemistry operating asymmetrically over different molecular enantiomers. However, we can also see some similarities between our proposed polarization-sensitive photochemistry and the traditional case of chiral molecular photochemistry. Our models and predictions are based on the realistic structures and conditions for the plasmonic nanoscale systems. Unlike the case of molecular systems, we show that optical CD, HE generation rates and rates for related photochemical processes should exhibit giant g-factors, or sensitivity to changes in the polarization of CPL. In the Au-dimers, the g-factor for hot electron generation can be as high as 15%, whereas the Ag-dimers exhibit even higher values for the chiral factor, up to ~ 60%. Such remarkable values should be certainly observable in photo-chemical experiments involving generation of HEs. For the broadband HE phenomena in the multi-NR structures, our results show a spectrally broad generation of hot electrons, with strong plasmonic CD for the wider spectral interval of 500 nm – 800 nm. Overall, our results offer interesting possibilities for new applications, such as polarization-sensitive photochemistry, asymmetric growth of chiral plasmonic nanostructures, or asymmetric photo-destruction of plasmonic complexes.

ASSOCIATED CONTENT

**Supporting Information**

This information includes the model used in COMSOL, the chiro-optical properties of Ag dimers, the structure of normal modes in the dimers and the maps of surface charge distribution.

AUTHOR INFORMATION

**Corresponding Author**

*E-mail: govorov@ohio.edu
*E-mail: jmwahng@gmail.com

ORCID

Alexander O. Govorov: 0000-0003-1316-6758
Tim Liedl: 0000-0002-0040-0173
Lucas V. Besteiro: 0000-0001-7356-7719

**Notes**

The authors declare no competing financial interest.

ACKNOWLEDGMENT

T. Liu and L.V.B was supported by the Institute of Fundamental and Frontier Sciences, University of Electronic Science and Technology of China. Z.W. was supported by National Basic Research Program of China (Project 2013CB933301) and National Natural Science Foundation of China (Project 51272038). This last author (A.O.G.) was funded via the 1000-talent Award of Sichuan, China. The collaborative Germany-US component was generously funded by the Volkswagen Foundation (T. Liedl and A.O.G.). Finally, A.O.G. holds Chang Jiang (Yangtze River) Chair Professorship in China.

(70) If we start the process of CPL illumination (LCP or RCP) over a racemic mixture of two chiral states of a NC (enantiomers: A-form and B-form), with time the NC mixture will depart from being racemic. We will instead have a system with A'- and B'-forms, where A' and B' are not enantiomers anymore, but new chiral states. The crystal volumes of the A' and B' forms will be different and their shapes will not be mirror-symmetric, since the total and local rates of HE generations exhibit a very strong CD. This case has not an obvious analogue in chiral molecular photochemistry, since we deal here with a fundamentally-different system – chiral plasmonic NCs. This fundamental difference stems, of course, from the fact that our system has a very large number of atoms, as compared to molecular systems.

# Supplementary Information

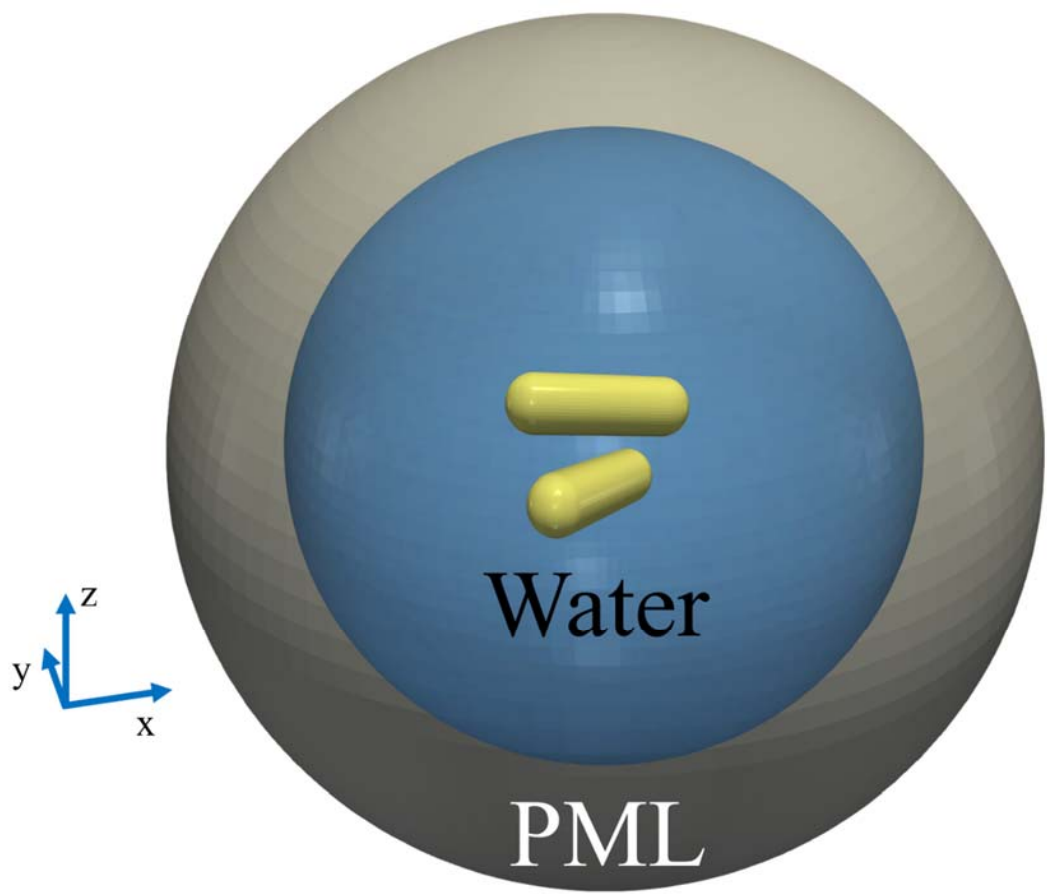

**Figure S1.** The COMSOL model used in the calculations. A plasmonic dimer (colored here as gold) was designed and its center of mass placed at the origin of coordinates, immersed in a 500 nm-radius water sphere (blue) and surrounded by a 200 nm-thick perfectly matched layer (PML, gray) that suppresses reflections at the boundaries and thus recreate the conditions of an isolated dimer. In the calculation, for the dimer, identically tetrahedral meshes were utilized for both plasmonic nanorods. To obtain the averaged extinction and generation rates of hot electrons, incident circularly polarized light ~~was sent~~ in orthogonal directions corresponding to the cartesian axes (x, y, and z). Then, the results were averaged for the three directions of incidence.

**Spectra for the Ag NR dimers.** It is interesting to compare dimers composed of different materials, and here we add data for Ag plasmonic NRs to complement the data for gold NCs presented in the main text. Experimentally, Au-NRs are widely available, whereas Ag-NRs are less common and comparatively harder to fabricate through wet-chemistry methods. As we pointed out in the main text, the Ag dimers showcase a larger Rabi splitting, as one does expect from the generally stronger and sharper plasmonic resonances present in good-quality silver.

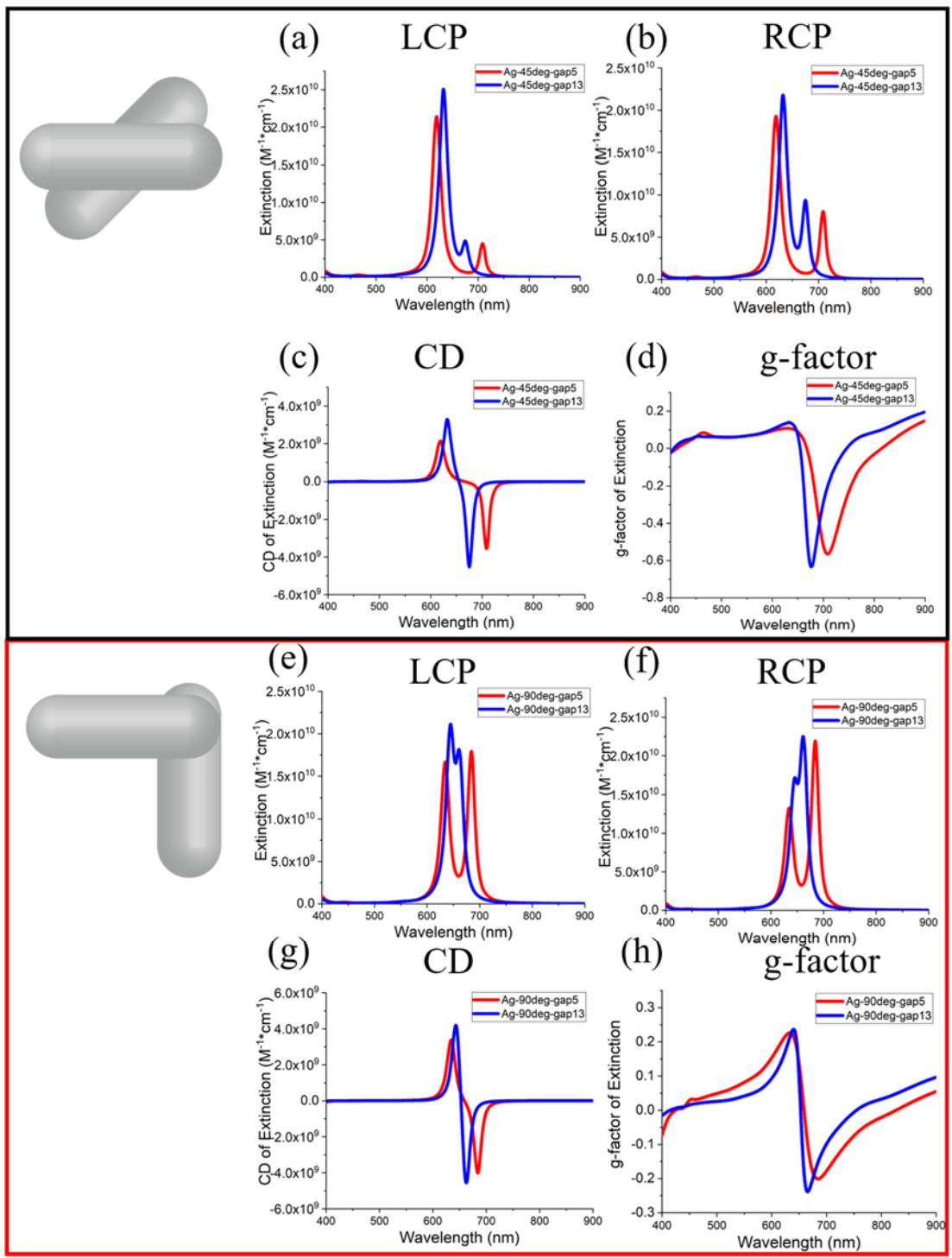

**Figure S2.** Chiro-optical data for the Ag-NRs. Calculated chiral extinction spectra for plasmonic dimers with 45° rotation angle (a-d), and 90° rotation angle (e-h). The optical data include the extinction spectra for LCP and RCP light, the CD spectra and the g-factor. These calculations were performed for the Ag-dimers with 5 nm and 13 nm gap sizes. The legend includes information about material, rotation angle of NRs and gap size.

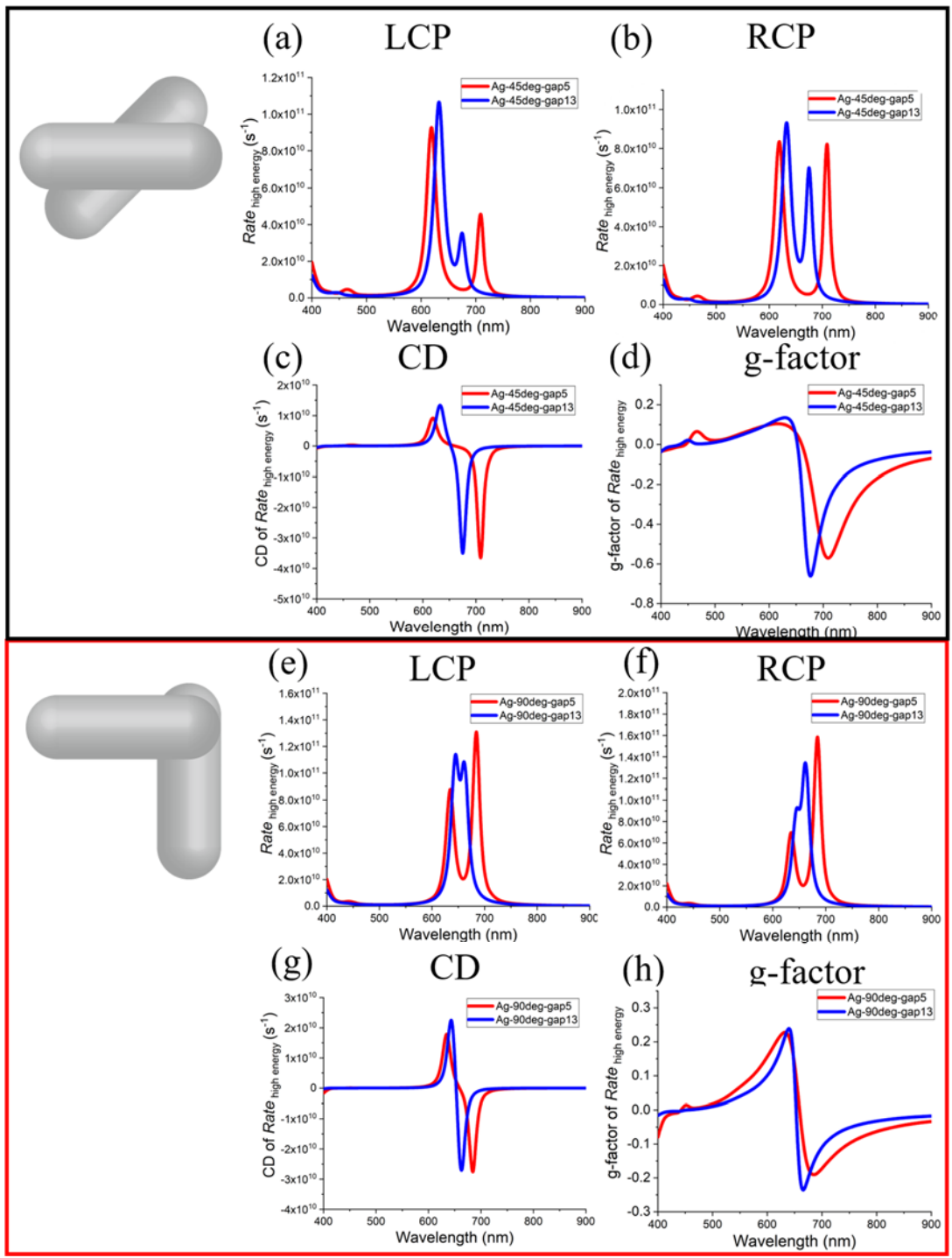

**Figure S3.** Optoelectronic (HE generation) data for the Ag-NRs under CPL. Calculated chiral spectra for plasmonic dimers with 45° rotation angle (a-d), and 90° rotation angle (e-h). The data include the HE generation rate spectra for LCP and RCP light, as well as the related CD spectra and g-factor. These calculations were performed for the Ag-dimers with 5 nm and 13 nm gap sizes. The legend includes information about material, rotation angle of NRs and gap size.

**Mode intensity and spectral asymmetry in the Rabi-splitting regime.** In the extinction spectra in Figure 2 (see main text), we clearly observe that the 45° NR structure produces strongly asymmetric spectra, whereas the 90° NR dimer exhibit a symmetric Rabi-split spectrum. We can easily understand this interesting difference by looking at the total plasmonic dipolar moment in our structures at the resonances. Figure S4(a) below shows that the 45° dimer has one weak and one strong dipolar mode, while the 90° dimer has two equally strong bright modes. Figure S4(b) shows the Rabi-splitting magnitudes as a function of the gap size between the two NRs for different materials (Au and Ag) and rotation angles.

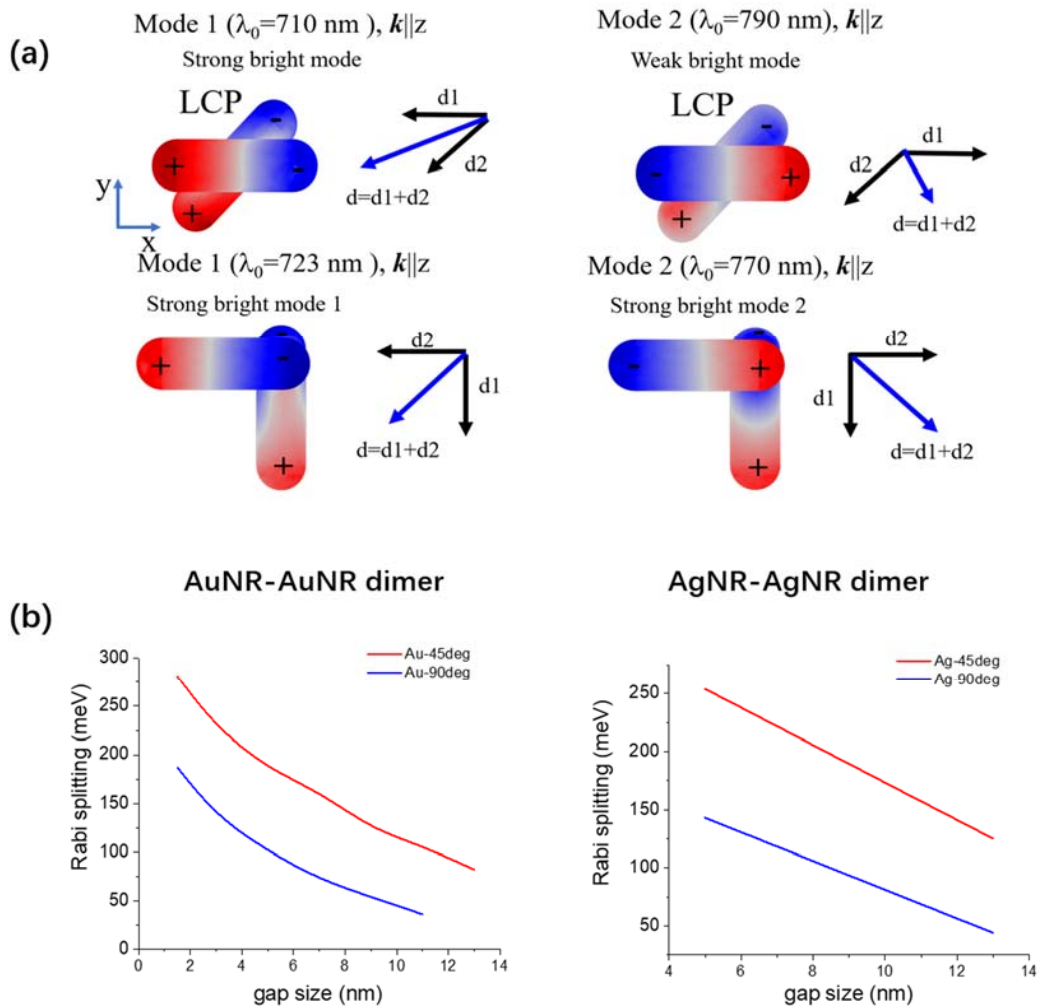

**Figure S4:** (a) Optical dipoles derived from the COMSOL-computed charge-density maps for both chiral geometries at the two plasmon peaks. (b) Rabi-splitting, expressed as the spectral distance

between the peaks in units on energy, for different NR-NR complexes as a function of the NR-NR gap.

**Comparison of the extinction-CD and HE-CD spectra for the short-wavelength interval of 400-600nm.** When we deal with the strong transversal resonance, the CD and HE-CD spectra behave similarly, as expected, but at the region of interband transitions in Au the difference between extinction-CD and HE-CD should be strong. Indeed, we observe this behavior in Figure S5. The CD g-factor is mostly positive for in the interval 400-500nm, whereas the HE-CD g-factor is negative in the same interval of 400-500nm. The physical reason for such difference between the optical and HE CD and g-factor spectra is that the extinction CD in the interval of 400-500nm in gold has a strong contribution from the interband absorption. Mathematically, absorption by a volume $\Delta V$ inside a NC is $\Delta Q_{abs} \sim \mathrm{Im}\, \varepsilon_{Au} |E_\omega|^2 \Delta V$. The case of HE-CD is different. The HE-CD depends on the electric field directly and the corresponding dissipation, $\Delta Q_{HE} \sim |E_{\omega,normal}|^2 \Delta S$, does not include the wavelength-dependent factor $\mathrm{Im}\, \varepsilon_{Au}$.

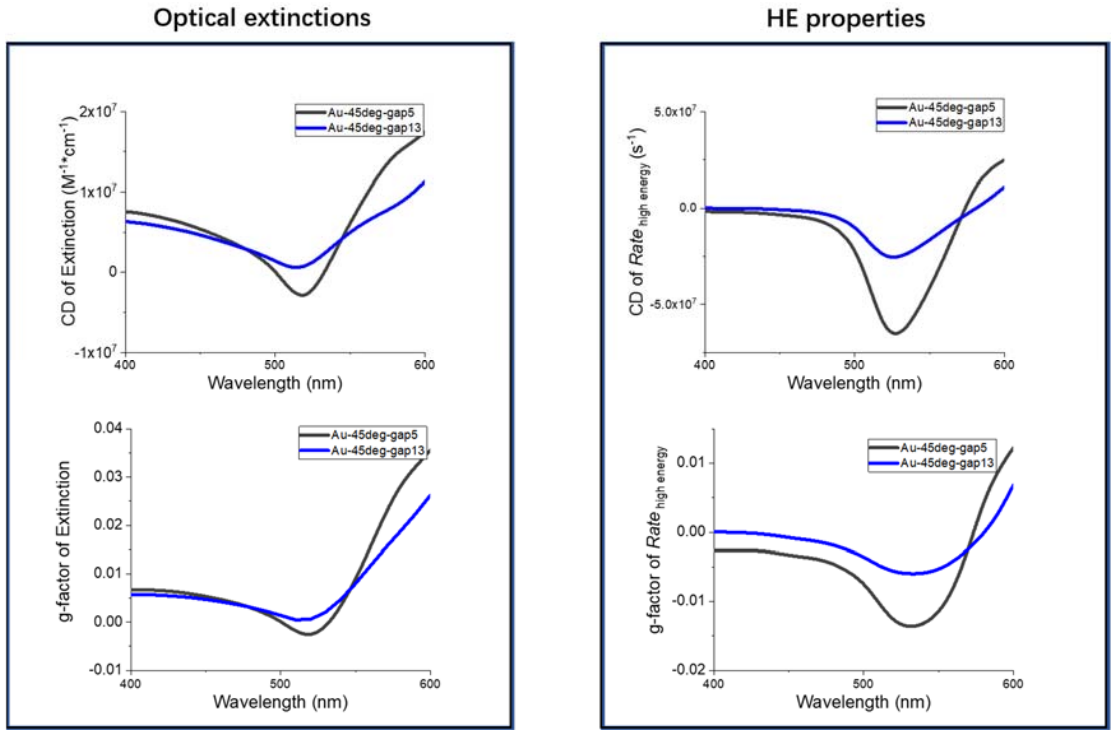

**Figure S5:** Chiroptical spectra of the 45°-dimer for the short-wavelength interval. We see a strong difference between the g-factors for extinction-CD and HE-CD.

Therefore, the wavelength dependencies are expected to be behave differently, as seen in Figure S5.

**Pulsed excitation regime.** Our formalism (Eqs.4,5 in the main text) can also be used for the pulsed regime of excitation. The average number of high-energy (over the barrier) HEs during a short pulse can be evaluated as: [S1]

$$N_{\text{high energy},\alpha} = Rate_{\text{high energy},\alpha} \cdot \overline{\tau}_{ee}(1-\frac{1-e^{-\Delta t/\overline{\tau}_{ee}}}{\Delta t/\overline{\tau}_{ee}}), \quad \alpha = L, R \quad (S1)$$

where $\Delta t$ is the pulse duration. To obtain Eq. S1, we used a rectangular pulse. In Eq. S1, $\overline{\tau}_{ee}$ is the average electron-electron scattering lifetime of high energy HEs, which can be derived from the Fermi liquid theory:

$$\overline{\tau}_{ee} = \frac{4\tau_0 E_F^2}{(\hbar\omega)^2} \quad (S2)$$

$$\tau_0 = \frac{128}{\pi^2\sqrt{3}}\omega_p^{-1} = \frac{128}{\pi^2\sqrt{3}}\sqrt{\frac{m^*\varepsilon_0}{ne^2}}_{(q\approx 0)}$$

where $\tau_0$ is the intrinsic electron-electron scattering lifetime in bulk metal, $\omega_p$ is the plasma frequency, which is associated with the effective electron mass of bulk metal $m^*$, the permittivity of free space is represented as $\varepsilon_0$, $n$ is the electron density, and $e$ is the elementary charge of electron. Note that we consider $\omega_p$ in the approximation of long wavelength ($q\approx 0$, with $q$ being the wavevector).

The estimate (Eqs. S2) for the average electron-electron (e-e) scattering lifetime ($\overline{\tau}_{ee}$) is strongly energy-dependent. In the visible and NIR wavelength ranges, the typical value of $\overline{\tau}_{ee}$ is ~10 fs, leading to $\overline{\tau}_{ee} \ll \Delta t$ and $N_{\text{high energy}} \approx \overline{\tau}_{ee} \cdot Rate_{\text{high energy}}$, obtained from Eq. S1. Qualitatively, longer wavelengths are more favorable to obtain larger values of $N_{\text{high energy},\alpha}$ because of the involved frequency dependences: $Rate_{\text{high energy}} \sim (\hbar\omega)^{-4}$ and $\overline{\tau}_{ee} \propto (\hbar\omega)^{-2}$.[S1,S2]

Then, the corresponding CD value for the average number of generated carriers is defined as

$$\Delta N_{\text{high energy}} = N_{\text{high energy},L} - N_{\text{high energy},R}$$

Although the calculations in the main text were done for steady-state generation and injection of HEs, experiments in the pulsed regime are also very popular and relevant to understand the dynamics of excited photocarriers.[S1,S2] Here we evaluate the average number of high energetic HEs ($N_{\text{high energy}}$) and the related CD ($\Delta N_{\text{high energy}}$) during an ultrashort pulse with a duration of 80 fs. As a model system, we use again the Au-dimer with 45° rotation angle and a 5 nm gap. Our calculations are based on Eqs. S1-S2 and shown below in Figure S6. More details for such time-dependent formalism can be found in Ref. S1. Figure S6 shows the results. As it is also the case for the CW illumination regime, we observe a CD effect in the optoelectronic responses. The CD signal for the average number of HEs during the pulse time interval is shown in Figure S6b.

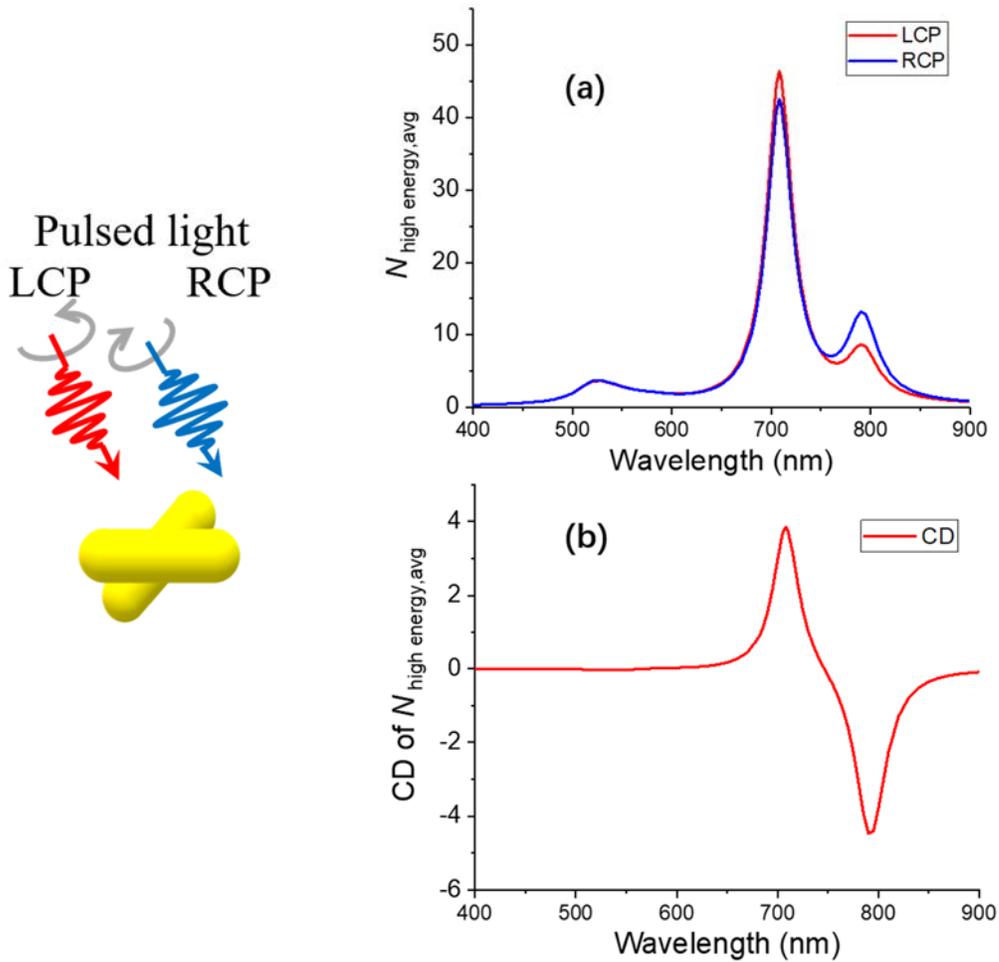

**Figure S6.** (a) Average number of generated hot electrons during the 80-fs pulse for LCP light (red) and RCP light (blue), respectively. (b) CD of the average number of generated hot electrons during the pulse. We show the case of Au-dimers with 45° rotation angle and 5 nm gap size. The flux during the pulse in this calculation was $2.5 \times 10^8$ W/cm$^2$, like in the experiment of Ref. S2.